\documentclass[12pt]{article} 
\def\bit{\begin{itemize}}
\def\eit{\end{itemize}}
\def\ben{\begin{enumerate}}
\def\een{\end{enumerate}}

\begin{document}

\begin{center}
  {\large\bf V.N. Gribov \\ 1930--1997} 
\end{center}

This is not an obituary\footnote{A shortened version of this text has
  been published in the January 1998 issue of ``Physics World''.}. 

This note is intended to make the physics world aware of the loss it
suffered on August 13 when professor Vladimir Gribov passed away, 
all of a sudden, 
in Budapest where he was steadily recovering after a mild stroke.

Vladimir Naumovich for youngsters, Volodya for friends, 
BH (his Russian initials taken as Latin letters)  
for colleagues world-wide. 

His devotion to physics was so intense, his knowledge, shared with anyone 
willing and prepared to listen, was so deep, 
that I feel one can still seek his advice, discuss problems with him,
trying to probe new ideas against the incredible physical intuition 
of this man, to match them with his ``picture''.
I am sure that many physicists, 
from St.Petersburg, Moscow and Novosibirsk, 
as well as those western theorists who knew him well, share this feeling. 

Gribov graduated from Leningrad University in 1952 when 
for a young man with Jewish blood
there was not a slightest chance to get a decent job. 
After Stalin was gone, the paranoid antisemitic wave receded. 
With the help of Ilya Shmushkevich and Karen Ter-Martirosyan,
Gribov, having served his term as a teacher at an evening school
for adults, was able to start his scientific career in Russia's first 
research institution --- the Physico-Technical Institute 
(later, Ioffe PTI) 
in Leningrad. 
Soon he was recognised as an informal leader of the 
theory group created and cherished by Shmushkevich. 
This group, under Gribov's lead, was to become one of the 
centres where the world-class physics of the 60's-70's was being
developed, later to be known as the ``Leningrad school''. 
In 1971 the theory group became a part of a new Nuclear Physics Institute 
(LNPI) in Gatchina, near Leningrad. 

In the late 50's Gribov was brought to Moscow and introduced to Lev Landau. 
It did not take long for Dau to form a high opinion of Gribov.
A special fund was created to allow a young physicist  
to commute from Leningrad  (400 miles one way) 
to participate in the weekly Moscow Landau seminars. 
 
There Volodya was to meet Isaak Pomeranchuk who became his close friend
and collaborator 
and made a great impact on Gribov the physicist.
  Gribov always referred to Pomeranchuk as his true Teacher.   
  He admired Chuk's intuition, style of doing research and 
his attitude to life and to physics. 

BH belonged to a generation of physicists, now almost extinct, for 
whom physics, in all its variety and complexity, was still felt as 
a single subject, who ``had a picture'', in his words. 
``He has a picture'', was Gribov's highest compliment, a universal 
formula ranging from appreciation to admiration.

Gribov was always open to discussion. 
He never refused, as far as I know, to discuss a physical problem,
be it of nuclear physics or elementary particle physics, 
cosmology or radiophysics, solid state physics or atomic physics. 
Not only did he know 
quantum physics as deeply as one can know it, 
he {\em felt}\/ quantum mechanics, he {\em thought}\/ quantum-mechanically. 
FSU physicists remember Yakov Zeldovich saying at the plenary session of
the annual Academy meeting: ``What a fool I was not to listen to what 
Volodya Gribov was telling me, long before Steven Hawking's 
work, on why and how black holes should radiate via quantum tunneling''.  
Gribov was the first to interpret an instanton --- a classical solution of 
non-linear Yang-Mills equations, found by Polyakov and collaborators ---
as an under-the-barrier trajectory linking vacua with different 
topology of the non-abelian field. 
This interpretation has become a common wisdom. 
He also came to the conclusion that classical fields 
(instantons, monopoles, etc.) are of little relevance 
for the long-standing problem of QCD colour confinement 
(which wisdom still awaits acceptance by the community).

``I am not smarter'', BH used to say, ``I just think more''...

\vspace{1 cm}

\begin{center}
  * * * 
\end{center}

For decades he was not allowed to travel abroad: 
a free-minded person was not the KGB's idea of a loyal citizen.
One can only guess how much harm Gribov's isolation has done 
to theoretical physics. 
Given an ever-red traffic-light on the road from LNPI to the West, 
many western physicists visited Leningrad in the 60's-70's  
to discuss new ideas with BH and his colleagues, to learn, and to   
go through the beneficial ordeal of a notorious ``Gribov seminar''. 

This was a legendary seminar. It had no time limit and would go on
as long as was necessary to establish the truth. 
Some visitors hated it and would never repeat that most dreadful experience
of their life; others loved it: finding the truth of the matter  
was at stake and the speaker would be the first to benefit.

For a speaker it was a test of self-confidence, of the depth of his or
her knowledge of the subject. 
Equally was it a challenge for the audience: to participate
during seminars (``to work on seminars'') 
was one of the two unquestionable duties of the members 
of the Gribov theory department. 
(The second one being: ``never refuse help to an experimentalist''.)

To understand the spirit of the seminar you have to accept the notion of 
``aggressive friendliness''. 
No merits counted, no excuses were given: 
a newcomer and a renouned academician were treated equally, 
that is equally 
amicably and aggressively. 
After 5 minutes of a smooth introduction 
BH
%% Gribov 
would jump to the 
blackboard and make his three points: what this guy is trying to tell us,
why it is ``all wrong'', and how the problem had to be approached. 
This would trigger a hot discussion involving all the audience
(including the speaker; 
though, markedly, there were historical exceptions when a speaker would 
leave the seminar room).

BH as a speaker would be given the same friendly treatment.
That is, the story goes, how 
Lev Lipatov, now a world-known theoretician and an academician, 
became a co-author of the famous 
Gribov \&\ Lipatov work of 1970-71 which laid the basis of a  
field-theoretical description of deep inelastic 
scattering and $e^+e^-$-annihilation. 
Gribov was presenting his work, 
and the young man made a couple of ``killing'' comments. 
Gribov got stuck trying to answer Lipatov's questions: ``Lev, you are a 
co-author already, help me'',  was the solution. 
 
Many a difficult problem was cracked in this fashion, at the blackboard,
in the noisy (and, in early days, smoky) atmosphere of PTI/LNPI seminars.

\vspace{1 cm}

\begin{center}
  * * * 
\end{center}

Gribov was never an icon, and a rosy picture of this character would be
unphysical and therefore false. He had a strong personality, strong
both in its rights and wrongs. 

It was not easy, to say the least, to argue with BH. 
In spite of his mind being fast, flexible and receptive, a prejudice
of his could be stone solid. 
You would not dare to start arguing with him before making absolutely clear
for yourself that the man was wrong. 
Such a dispute could eventually rise to a fight, sometimes reaching heights
which any socium with a minimal awareness of good 
manners would classify as absolutely unacceptable. 
To shout at your boss, though, was pretty safe: 
Gribov and his Leningrad colleagues  
always remembered the heritage of Ilya Shmushkevich: ``a scientific argument
cannot lead to administrative conclusions'' (sounds much better in the
original Soviet newspeak). 

Neither was Gribov always right in his vision. It took a good 10 years
for him to accept quarks as the true basis of hadron physics. 
He encouraged, though, his young students to play with a new
hypothesis and discussed
with them applications of the quark picture to hadron scattering. 
Hence, the famous Frankfurt-Levin ratio of the pion-proton and
proton-proton cross sections of the early days of the quark model.

\vspace{1 cm}

\begin{center}
  * * * 
\end{center}

``When I was young I was happy to see the pieces of a lengthy  
  calculation cancel and produce a zero result. This told me that
  I had been smart and had not make a mistake. 
  Only later did I recognise that
  this was stupid: a good physicist should know a priori 
  that the answer will be zero.''
This recollection of Gribov's can tell you much about his research style, 
a very special way of attacking a difficult theoretical problem that he  
developed and used with brilliance. He had a profound knowledge and skill
in using mathematical methods in physics. However, describing his results
Gribov would not stress the mathematical difficulty, 
not even the mathematical beauty, of the solution he found. 
What mattered most was, once again, ``a picture''. He would approach the 
problem from different angles, abstracting its essential features 
and illustrating them with the help of simplified models and analogues from
different branches of physics, solid state physics being his favourite source
of inspiration. 

You were led to see that the answer is correct because there is a clear 
physical picture behind its structure and its properties, not merely because
it has emerged as a result of a derivation following the mathematical 
deduction rules. 
People unfamiliar with this style were often confused.  After Gribov's talk  
some felt they were being cheated:  
a couple of chalk drawings, 
a strain of hand-waving arguments, and --- here you are: that's the answer?
Such listeners were not aware that they fell victim of the speaker's 
generosity:
for Gribov it went without saying that the receiving party is capable 
of reproducing the necessary mathematical calculations and analysis, 
this being a default professional quality. He was talking physics.
Even when a mathematical framework to envelope a foreseen physical answer 
was not developed, and thus the problem not solved, 
this would not stop him from sharing his ideas and 
arguments with anyone willing to listen. 
Physics was given top priority, ambitions put aside. 
``Physics goes first'' was the motto.
  
One inside story to illustrate the point.
A project BH was pursuing with his student had run at some point into
a rather difficult mathematical problem. 
The student was given a page of notes where the basic idea 
of how to approach the problem was briefly explained, followed by few lines 
of calculations. 
He was shocked to find out that the very first equation that the
ma\^{\i}tre 
had written was wrong. Having done the job and having noticed that 
the other nine equations that followed were dead wrong as well, the student
arrived at the answer.  
He compared it with what was written on the bottom of the Gribov note,
and the answer there was the correct one.   
Weird though it might seem, it was neither a miracle nor an accident.
According to Alexei Anselm, for many years Gribov's collaborator and 
friend, --- ``Working with BH you had a strange feeling 
that numbers were his personal friends: 
all those factors of $2$ and $\pi$ simply knew their place in
Gribov's formulae''. 

\vspace{1 cm}

\begin{center}
  * * * 
\end{center}

Gribov left Leningrad in 1980, on the eve of turning 50. It was a hard
blow for the LNPI theory laboratory --- the Gribov laboratory. It
remained a group of top-class theorists but was never again the unique
team that it had been. 
The loss for BH proved to be comparable, if
not stronger. Having moved to Moscow for personal reasons, 
he found himself pretty much in isolation. 
The Landau Institute for Theoretical Physics in Chernogolovka, with
which he formally became affiliated, 
had its established orderly way of things.
It goes without saying that everyone respected Gribov, 
a ``ring-bearer'' of the Landau tradition. 
At the same time, the community as a whole
was not ready to accommodate such a disturbing and virulent force: he
did not fit into the style of Chernogolovka seminars. 

Later he lived permanently in Budapest with his new family and, in a
wider world, was being warmly received in the US and Sweden, France
and Italy. 
Recently Gribov, 
as a Humboldt awardee,
enjoyed the hospitality of the Nuclear Physics Institute in Bonn. 
However, no place was to be found in the West for a man about to turn
60, where he could start a new school and work in a team --- a
natural Gribov environment.

Many a year went under the strain of personal tragedy.
Lenya Gribov, the son of Volodya and his first wife Lilya Dubinskaya,  
perished in a mountaineering accident a few months after defending 
his Ph.D. in theoretical physics.
Volodya kept cursing himself for having infected Lenya with his 
passion for mountains. 
Neither time nor the loving attention of wife Julia
and step-son Palik could help to heal the wound. 

When asked by Julia what physics meant for him, 
Volodya said that 
he had realised quite early that if he made an effort he could find the truth. 
And therefore, he had decided, he must.
 He kept working,  
 working on the most challenging problem, 
 working with unmatched persistence and intensity which has only
 doubled after the loss of his son.

Being a perfectionist, BH would not write a paper before he had the final
solution of the entire problem that he had set for himself.   
August 13, 1997 caught Volodya Gribov in the process of writing up the
work concluding his 20-year-long study of the problem of quark confinement
in quantum chromodynamics. 

\vspace{1 cm}

\begin{center}
  * * * 
\end{center}

Gribov's contribution to physics deserves a special study.
It suffices to say that his name is attached to many a key notion of 
modern theoretical physics: 
Gribov-Froissart projection and the Gribov vacuum pole (Pomeron),
factorisation, Reggeon calculus, Gribov diffusion,
the AGK 
% Abramovsky-Gribov-Kancheli 
cutting rules, the Gribov bremsstrahlung theorem, 
Gribov-Lipatov evolution equations, 
and many more.

Gribov's impact on modern physics is deeper than it is known to be.

One of his jewels    
  ``Interaction of photons and electrons with nuclei at high energies'' 
%% (1969)
where the space-time picture of particle interactions at
high energies was established, found its way through the iron curtain.  
The key elements of this work were 
incorporated into the famous Feynman
book which laid the foundation of the parton model. 
The Feynman-Gribov parton model, that is. 

Gribov with Alexander Migdal developed 
%% in 1968 
an ingenious technique for 
ana\-lysing dynamical systems with long-range fluctuations,
which triggered a breakthrough in solid state physics. 
The physics of solids near the critical temperature 
proved to be similar to that of the so-called strong-coupling regime 
of high-energy hadron-hadron interactions. 
The subsequent works of the ``two Sashas'' 
--- Polyakov and Migdal --- and a contemporary more general 
treatment suggested by L.~Kadanoff and K.~Wilson have established 
the scaling solution of the problem of the  
second order phase transitions. 

Gribov's QCD studies produced a brilliant physical explanation of
asymptotic freedom, based on an early observation of the
anti-screening phenomenon made by Julij Khriplovich in a pre-historic 1969.
In 1977 Gribov demonstrated the inconsistency of the standard 
field-theoretical treatment of gluon fields 
(Gribov copies, the Gribov horizon). 
Later he suggested the quark confinement scenario based on
super-critical binding of light quarks by a quasi-Coulomb colour interaction.

His last works remain to be discovered, understood and developed.

\vspace{1 cm}

\begin{center}
  * * * 
\end{center}

Vladimir Gribov believed in the Truth in physics. 
Not that he was a na\"{\i}ve man, but he could not (or rather did not want to)
understand how some people calling themselves physicists would
politely listen to and applaud ``nonsense''.
He thought that everyone shares his ``physics goes first'' belief  and is
ready to put aside any political, mercantile considerations
when a physical issue is at stake. 
In our pragmatic world such a scenario does not look very realistic.
However, since his commitment to physics was close to religious, 
we can consider it as Gribov's prophecy
for the physics world of the future.

\vfill
\flushright
{Yuri Dokshitzer}

\end{document}